\documentclass[onecolumn]{aastex63}
\usepackage{amsmath,amsfonts,amssymb}
\usepackage{graphicx,subfigure,multirow}

\begin{document}
\title{PSR J0030+0451, GW170817, and the nuclear data: joint constraints on equation of state and bulk properties of neutron stars}
\author{Jin-Liang Jiang}
\author{Shao-Peng Tang}
\affil{Key Laboratory of Dark Matter and Space Astronomy, Purple Mountain Observatory, Chinese Academy of Sciences, Nanjing 210023, China.}
\affil{School of Astronomy and Space Science, University of Science and Technology of China, Hefei, Anhui 230026, China.}
\author{Yuan-Zhu Wang}
\affil{Key Laboratory of Dark Matter and Space Astronomy, Purple Mountain Observatory, Chinese Academy of Sciences, Nanjing 210023, China.}
\author{Yi-Zhong Fan}
\author{Da-Ming Wei}
\affil{Key Laboratory of Dark Matter and Space Astronomy, Purple Mountain Observatory, Chinese Academy of Sciences, Nanjing 210023, China.}
\affil{School of Astronomy and Space Science, University of Science and Technology of China, Hefei, Anhui 230026, China.}
\email{Corresponding author: yzfan@pmo.ac.cn}

\begin{abstract}
Very recently the \emph{NICER} collaboration has published the first-ever accurate measurement of mass and radius together for \objectname{PSR J0030+0451}, a nearby isolated quickly rotating neutron star (NS). In this work we set the joint constraints on the equation of state (EoS) and some bulk properties of NSs with the data of \objectname{PSR J0030+0451}, GW170817, and some nuclear experiments. The piecewise polytropic expansion method and the spectral decomposition method have been adopted to parameterize the EoS. The resulting constraints are consistent with each other. Assuming the maximal gravitational mass of nonrotating NS $M_{\rm TOV}$ lies between $2.04 M_{\odot}$ and $2.4 M_{\odot}$, with the piecewise method the pressure at twice nuclear saturation density is measured to be $3.19^{+2.63}_{-1.35}\times 10^{34}~{\rm dyn~cm^{-2}}$ at the $90\%$ level. For an NS with canonical mass of $1.4 M_\odot$, we have the moment of inertia $I_{1.4} = {1.43}_{-0.13}^{+0.30} \times 10^{38}~{\rm kg \cdot m^2}$, tidal deformability $\Lambda_{1.4} = {370}_{-130}^{+360} $, radius $R_{1.4}= {12.1}_{-0.8}^{+1.2}~{\rm km}$, and binding energy $BE_{1.4} = {0.16}_{-0.02}^{+0.01} M_{\odot}$ at the $90\%$ level, which are improved in comparison to the constraints with the sole data of GW170817. These conclusions are drawn for the mass/radius measurements of \objectname{PSR J0030+0451} by \citet{2019arXiv191205702R}. For the measurements of \citet{2019arXiv191205705M}, the results are rather similar.
\end{abstract}

\section{Introduction}

As the most compact directly observable objects in the universe, neutron stars (NSs) are made of material mainly at supranuclear densities. The equation of state (EoS) of NSs, i.e., the relation between pressure and energy density, describes the general properties of such dense matter. Numerous theoretical EoSs have been developed in the literature \citep[see][for a review]{2017RvMP...89a5007O} and the observational data are highly needed to distinguish between them.

The masses and radii of the NSs, the nuclear experiment data, and the gravitational-wave data of neutron star mergers are widely known as the powerful probes \citep[see][for comprehensive reviews]{2012ARNPS..62..485L,2016ARA&A..54..401O,2019PrPNP.10903714B}. As witnessed in these two years, the discovery of the binary neutron star merger event GW170817 in the O2 run of advanced LIGO/Virgo detectors \citep{2017PhRvL.119p1101A}, together with some reasonable assumptions and empirical relationships, have significantly boosted research on the EoS of neutron stars \citep[e.g.,][]{2018PhRvL.121p1101A,2018PhRvL.120q2703A,2018PhRvL.120z1103M,2018PhRvL.121i1102D,2019ApJ...885...39J}. The masses of a small fraction of Galactic NSs in the binary systems have been precisely measured and the record of the most massive NS was broken over and over. The latest record is held by \objectname{PSR J0740+6620}, a millisecond pulsar with a mass of $2.14^{+0.10}_{-0.09} \rm M_{\odot}$ \citep{2019NatAs.tmp..439C}. Much more massive NS might exist, as indicated in the updated constraints on the mass of \objectname{PSR J1748-2021B} \citep[][by assuming random inclinations a median pulsar mass of $2.548_{-0.078}^{+0.047} \rm M_\odot$ is inferred with the 11 years of continued observation data of the Green Bank Telescope]{Clifford2019}. The radii of some NSs have been previously inferred in a few ways \citep[see][for a recent review]{2016ARA&A..54..401O}. However, these results are much more model dependent than the masses because the radius ``measurements" are usually indirect and suffer from some systematic uncertainties, including for instance the composition of the atmosphere, the distance of the source, the interstellar extinction, and the brightness \citep[see][for the detailed discussion]{2016EPJA...52...63M}. Pulse Profile Modeling \citep[also known as waveform modeling;][]{2014ApJ...792...87P} exploits the effects of General and Special Relativity on rotationally modulated emission from neutron star surface hot spots \citep[see][for a review]{2016RvMP...88b1001W} and does not suffer from these limitations. In principle, the Pulse Profile Modeling can deliver simultaneous measurements of mass and radius at an unprecedented level of a few percent \citep[e.g.,][]{2014ApJ...787..136P,2016ApJ...832...92O}. Such precise measurements are the primary scientific goal of \emph{Neutron Star Interior Composition Explorer} \citep[\emph{NICER};][]{2016SPIE.9905E..1HG}, a pioneering soft X-ray telescope installed on the International Space Station in 2017. Thanks to the successful performance of \emph{NICER}, very recently the first-ever accurate measurement of mass and radius together for \objectname{PSR J0030+0451}, a nearby isolated quicklyrotating NS, has been achieved \citep{2019arXiv191205702R,2019arXiv191205705M} and has far-reaching implications on the EoS of NSs \citep[e.g.][]{2019arXiv191205705M,2019arXiv191205703R,2018A&A...616A.105S}.

In this work, we aim to set the joint constraints on the EoS and some bulk properties of NSs with the data of \objectname{PSR J0030+0451}, GW170817, and some nuclear experiments. This work is organized as follows. We describe our parameterization methods, data sets, and models in Section \ref{sec:method}. The results of joint constraints on the EoS and some bulk properties of NSs are presented in Section \ref{sec:result}. We summarize our conclusion with a discussion in Section \ref{sec:conc_disc}.

\section{Method}
\label{sec:method}

\subsection{Parameterizing Methods}
\label{sec:param_method}
Parameterized representations of the EoS can reasonably describe the main properties of the dense matters in NSs \citep[see][for a review]{2019PrPNP.10903714B}. Currently there are two widely adopted methods to parameterize the EoS, namely, the piecewise polytropic expansion method \citep{1999AIPC..493...60V,2009PhRvD..79l4032R,2015PhRvD..91d3002L} and the spectral decomposition method \citep{2010PhRvD..82j3011L,2018PhRvD..98f3004C}.

In this work we take the same piecewise polytropic expansion method and parameter settings used in \citet{2019ApJ...885...39J}, along with the same pressure-based spectral decomposition method used in \citet{2018PhRvL.120q2703A} to parameterize the EoS. The piecewise polytropic expansion method uses four pressures $\{P_{\rm 1n_s}, P_{\rm 1.85n_s}, P_{\rm 3.7n_s}, P_{\rm 7.4n_s}\}$ located, respectively, at four different rest mass densities $\{\rho_{\rm 1n_s}, \rho_{\rm 1.85n_s}, \rho_{\rm 3.7n_s}, \rho_{\rm 7.4n_s}\}$ to parameterize the EoS, where $n_{\rm s}$ is the nuclear saturation density. Between each pair of adjoining bounds of densities, we approximately take a polytropic form. While the spectral decomposition method uses four expansion parameters $\{\gamma_0, \gamma_1, \gamma_2, \gamma_3\}$ to describe the relation between the pressure and the adiabatic index, and thus uniquely determine an EoS.

\subsection{Constraints of the EoS}
\label{sec:cons_eos}

Nuclear experiments and gravitational-wave data have been proved to be essential in constraining the EoS \citep[e.g.,][]{2014EPJA...50...40L,2017ApJ...848..105T,2018PhRvL.120q2703A,2018PhRvL.120z1103M,2018PhRvL.121i1102D,2018PhRvL.121p1101A}. As summarized in Sec. 2.4 of \citet{2019ApJ...885...39J}, the current nuclear data have set interesting constraints on both $P_{\rm 1n_{\rm s}}$ and $P_{\rm 1.85n_{\rm s}}$, which are $4.70 \ge P_{\rm 1n_{\rm s}}/(10^{33} \rm{dyn~cm^{-2}}) \ge 3.12$ and $P_{\rm 1.85n_{\rm s}}/(10^{34} \rm dyn~cm^{-2}) \ge 1.21$, respectively. We also follow that paper (Sec. 2.5 therein) to take into account the constraints set by the gravitational-wave data of GW170817. However, in this work we do not consider the low-mass X-ray binary data further because the latest \emph{NICER} measurement of \objectname{PSR J0030+0451} is much more direct and suffers from significantly fewer systematic uncertainties.

The \emph{NICER} teams have measured the mass and the radius of \objectname{PSR J0030+0451} through pulse profile modeling methods. \citet{2019arXiv191205702R} reported a mass (radius) $1.34^{+0.15}_{-0.16} \rm M_{\odot}$ $(12.71^{+1.14}_{-1.19} \rm km)$, while \citet{2019arXiv191205705M} reported a mass (radius) $1.44^{+0.15}_{-0.14} \rm M_{\odot}$ $(13.02^{+1.24}_{-1.06} \rm km)$, which are well consistent with each other. We mimic the mass-radius posterior distribution of this source by a two-dimensional Gaussian kernel density estimation (KDE) to constrain the EoS:
\begin{equation}
    P(M, R) = KDE(M, R|\vec{S}),
    \label{eq:mr_post}
\end{equation}
where $\vec{S}$ is posterior samples taken from the best fitting three ovals case of \citet{miller_data} and the ST+PST case of \citet{riley_data}. Unless specified, we only show the results with the data of \citet{riley_data}, because the data of \citet{miller_data} yield rather similar results, as shown in Tab.\ref{tb:canonical}. Besides, we do not consider the impact of spin on the radius because \citet{2019arXiv191205703R} have shown that the spin effect of this source is small in comparison to the rather large systematic uncertainties.

In addition to satisfying the above bounds, the EoS should meet the following general requests: (i) the causality condition, i.e., the speed of sound of the dense matter can never exceed the speed of light $c$; (ii) the microscopical stability condition, i.e., the pressure cannot be smaller in denser matters; (iii) the maximal gravitational mass of nonrotating NSs ($M_{\rm TOV}$) should be above all those accurately measured. Note that for the rapidly rotating NSs, the gravitational mass has been enhanced. For \objectname{PSR J0740+6620}, the enhancement of the gravitational mass is about $0.01 M_\odot$ \citep{2016MNRAS.459..646B,2018ApJ...858...74M}. Thus we have a robust lower limit of $M_{\rm TOV} \ge 2.04 \rm{M_{\odot}}$ based on the mass measurement of \objectname{PSR J0740+6620} \citep{2019NatAs.tmp..439C}. Additionally, the adiabatic index $\Gamma(p)$ for the spectral decomposition method are limited in the range $[0.6, 4.5]$ \citep{2018PhRvL.121p1101A}.

\subsection{Models}

As mentioned in Section \ref{sec:param_method}, two parameterization methods are adopted in this work. To investigate the impact of $M_{\rm TOV}$ on constraining the EoS, we consider the possible regions of $(2.04, 2.40) {M_{\odot}}$ (i.e., the low $M_{\rm TOV}$ case) and $(2.40, 2.90) M_{\odot}$ (i.e., the high $M_{\rm TOV}$ case), respectively\footnote{The exact value of $M_{\rm TOV}$ is still unknown. In a recent study incorporating an explicitly isospin-dependent parametric EoS of the neutron star matter, $M_{\rm TOV}$ is found to be $\leq 2.40 M_\odot$ \citep{ZhangNB2019}. A higher $M_{\rm TOV}$, however, may have been suggested by the updated mass measurement/estimate of \objectname{PSR J1748-2021B} \citep{Clifford2019}. Therefore, in this work we investigate both the low and high $M_{\rm TOV}$ cases.}. With these considerations we carry out four tests:

(i) Using the piecewise method to parameterize the EoS, assuming $M_{\rm TOV} \in (2.04, 2.40) M_{\odot}$.

(ii) The same as (i), except assuming $M_{\rm TOV} \in (2.40, 2.90) M_{\odot}$.

(iii) Using the spectral method to parameterize the EoS, assuming the low $M_{\rm TOV}$ case.

(iv) The same as (iii), except assuming the high $M_{\rm TOV}$ case.

To investigate the role of prior in each test, we also construct the prior for piecewise method and spectral method, where we just take into account the causality condition, the microscopical stability condition, and the lower limit $M_{\rm TOV}>2.04 M_{\odot}$, while in the spectral case the adiabatic index is assumed to be within the range $[0.6, 4.5]$.

Since we combine the data of GW170817 and \objectname{PSR J0030+0451} to do the analyis, the likelihood in each test will take the form
\begin{equation}
    L \propto \exp{ [-2\int_0^{\infty} \frac{|\tilde{d}(f)-\tilde{h}(f; \vec{\theta}_{\rm gw})|^2}{S_n(f)}\, df]} \times P(M,R),
    \label{eq:likelihood}
\end{equation}
where $\vec{\theta}_{\rm gw} = \{\mathcal{M}_{\rm c}, q, \chi_1, \chi_2, \theta_{\rm jn}, t_{\rm c}, \Psi, \Lambda_1, \Lambda_2 \}$ are the gravitational-wave parameters. The $\mathcal{M}_{\rm c}$, $q$, $\chi_i$, $\theta_{\rm jn}$, $t_{\rm c}$, $\Psi$, and $\Lambda_i$ are chirp mass, mass ratio, spin of the $i$th neutron star, inclination angle, coalescence time, polarization, and tidal deformability of the $i$th neutron star, respectively. The $\tilde{d}(f)$, $\tilde{h}(f)$, and $S_n(f)$ are frequency domain gravitational-wave data of GW170817, frequency domain waveform, and power spectral density of the GW170817 data, respectively. The two tidal deformabilities of source NSs of GW170817 and the mass/radius of \objectname{PSR J0030+0451} are determined by
\begin{equation}
    \begin{aligned}
    \Lambda_1 & = \Lambda_1(\vec{\theta}_{\rm eos}, M_1),~~~ \Lambda_2 = \Lambda_2(\vec{\theta}_{\rm eos}, M_2), \\
    M & = M(\vec{\theta}_{\rm eos}, p_{\rm c}),~~~R = R(\vec{\theta}_{\rm eos}, p_{\rm c}),
    \label{eq:transform}
    \end{aligned}
\end{equation}
where $\vec{\theta}_{\rm eos}$ are parameters needed to describe an EoS, in the case of the piecewise method, $\vec{\theta}_{\rm eos} = \{P_{\rm 1n_s}, P_{\rm 1.85n_s}, P_{\rm 3.7n_s}, P_{\rm 7.4n_s} \}$, whereas in the case of spectral method, $\vec{\theta}_{\rm eos} = \{\gamma_0, \gamma_1, \gamma_2, \gamma_3\}$. The $p_{\rm c}$ is the central pressure of \objectname{PSR J0030+0451}. The $M_1$ and $M_2$ are evaluated from $\mathcal{M}_{\rm c}$ and $q$. Since we marginalize the distance and phase in analyzing the gravitational wave, there are only $12$ parameters in total, namely, $\vec{\theta} = \vec{\theta}_{\rm gw} \cup \vec{\theta}_{\rm eos} \cup p_c $, and we sample these parameters using the PyMultiNest code \citep{2014A&A...564A.125B,2016ascl.soft06005B} implemented in Bilby \citep{2019ApJS..241...27A,2019ascl.soft01011A}, while the likelihood contribution of gravitational-wave is calculated using PyCBC \citep{2018ascl.soft05030T,2019PASP..131b4503B}.

\begin{table}[]
\begin{ruledtabular}
\centering
\caption{Priors and Posteriors of $\vec{\theta}_{\rm eos}$ for Different Parameterizing Methods}
\label{tb:priors_and_post}
\begin{tabular}{ccccc}
Methods                      & Parameters                         & Prior distributions        &     $90\%$ Range (low $M_{TOV}$)   &     $90\%$ Range (high $M_{TOV}$) \\ \hline
\multirow{4}{*}{Spectral}    & $\gamma_0$                         & U\tablenotemark{\dag}(0.2, 2.0) &   $0.80_{-0.44}^{+0.49}$   &   $0.69_{-0.32}^{+0.42}$    \\
                             & $\gamma_1$                         & U(-1.6, 1.7)             &   $0.18_{-0.62}^{+0.68}$   &   $0.32_{-0.53}^{+0.51}$    \\
                             & $\gamma_2$                         & U(-0.6, 0.6)             &   $-0.02_{-0.20}^{+0.16}$   &   $-0.04_{-0.16}^{+0.14}$    \\
                             & $\gamma_3$                         & U(-0.02, 0.02)           &   $-0.00_{-0.01}^{+0.02}$   &   $0.00_{-0.01}^{+0.01}$    \\  \hline
\multirow{4}{*}{Piecewise}  & $P_{\rm 1n_s}/(10^{33}{\rm dyn\,cm^{-2}})$    & U(3.12, 4.7)  &   $3.96_{-0.72}^{+0.66}$   &   $3.93_{-0.72}^{+0.69}$    \\
                             & $P_{\rm 1.85n_s}/(10^{34}{\rm dyn\,cm^{-2}})$ & U(1.21, 8.0)  &   $2.42_{-1.12}^{+2.45}$   &   $2.85_{-1.51}^{+2.05}$    \\
                             & $P_{\rm 3.7n_s}/(10^{35}{\rm dyn\,cm^{-2}})$  & U(0.6, 7.0)   &   $2.90_{-0.88}^{+0.97}$   &   $5.15_{-1.41}^{+1.59}$    \\
                             & $P_{\rm 7.4n_s}/(10^{36}{\rm dyn\,cm^{-2}})$  & U(0.3, 4.0)   &   $2.01_{-1.36}^{+1.65}$   &   $2.30_{-1.49}^{+1.41}$    \\
\end{tabular}
\tablenotetext{\dag}{Uniform distribution}
\end{ruledtabular}
\end{table}

\section{Constraint Results}
\label{sec:result}

Following the implement of methods described in Sec.\ref{sec:param_method}, all the tests are carried out to get the posterior distributions of $\vec{\theta}$ using Bayesian parameter estimation with the gravitational-wave data, the $M-R$ data of \objectname{PSR J0030+0451}, as well as the constraints from nuclear experiments and the limits on $M_{\rm TOV}$. The $90\%$ uncertainties of the EoS parameters are summarized in Table \ref{tb:priors_and_post}. Besides, although the spectral method in all the cases yield slightly higher pressure, larger redshift, and smaller radius than the piecewise method given the same conditions, these two methods yield results highly consistent with each other (as shown in Fig.\ref{fig:eos}, Figure \ref{fig:canonical}-\ref{fig:bulk}, and Table \ref{tb:canonical}). It is also found that the harder region of EoS in the prior is largely excluded by the gravitational-wave data and the $M-R$ measurement results of \objectname{PSR J0030+0451}.

\subsection{The EoS and the Sound Speeds of Neutron Stars}

\begin{figure}[htbp]
    \subfigure[]{
    \begin{minipage}[t]{0.47\linewidth}
    \includegraphics[width=3.8in]{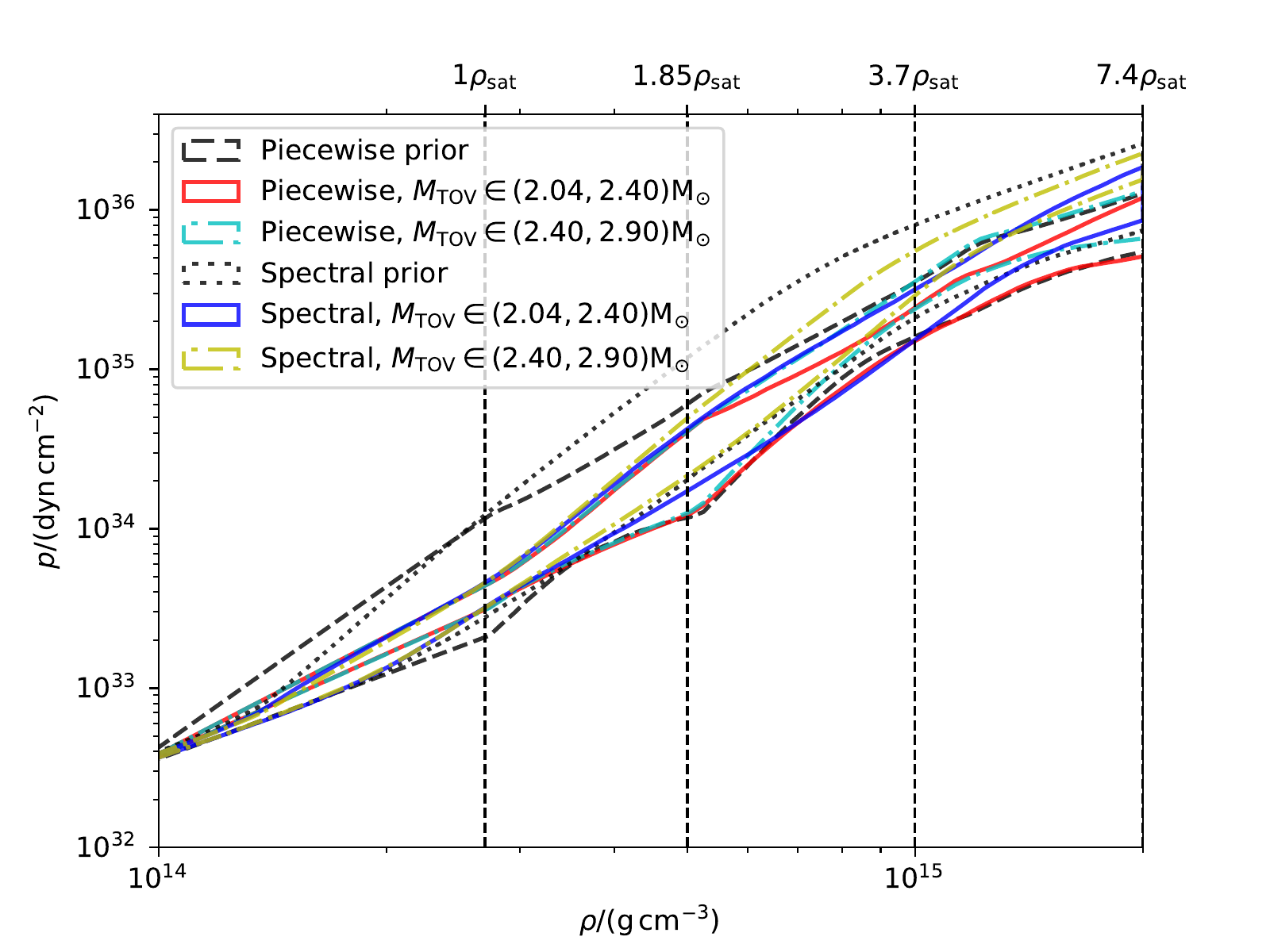}
    \end{minipage}
    }
    \subfigure[]{
    \begin{minipage}[t]{0.47\linewidth}
    \includegraphics[width=3.8in]{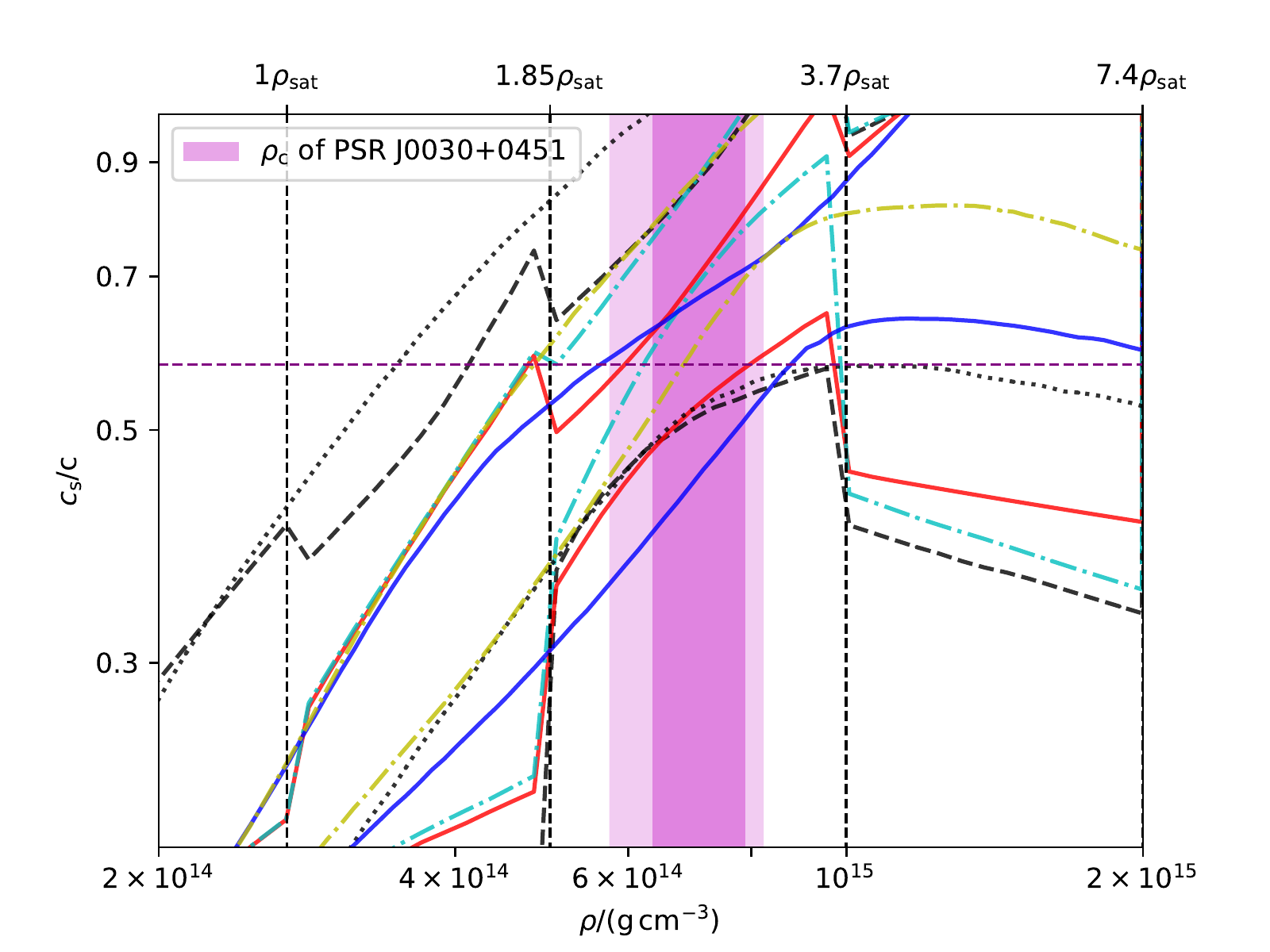}
    \end{minipage}
    }
    \subfigure[]{
    \begin{minipage}[t]{0.47\linewidth}
    \includegraphics[width=3.8in]{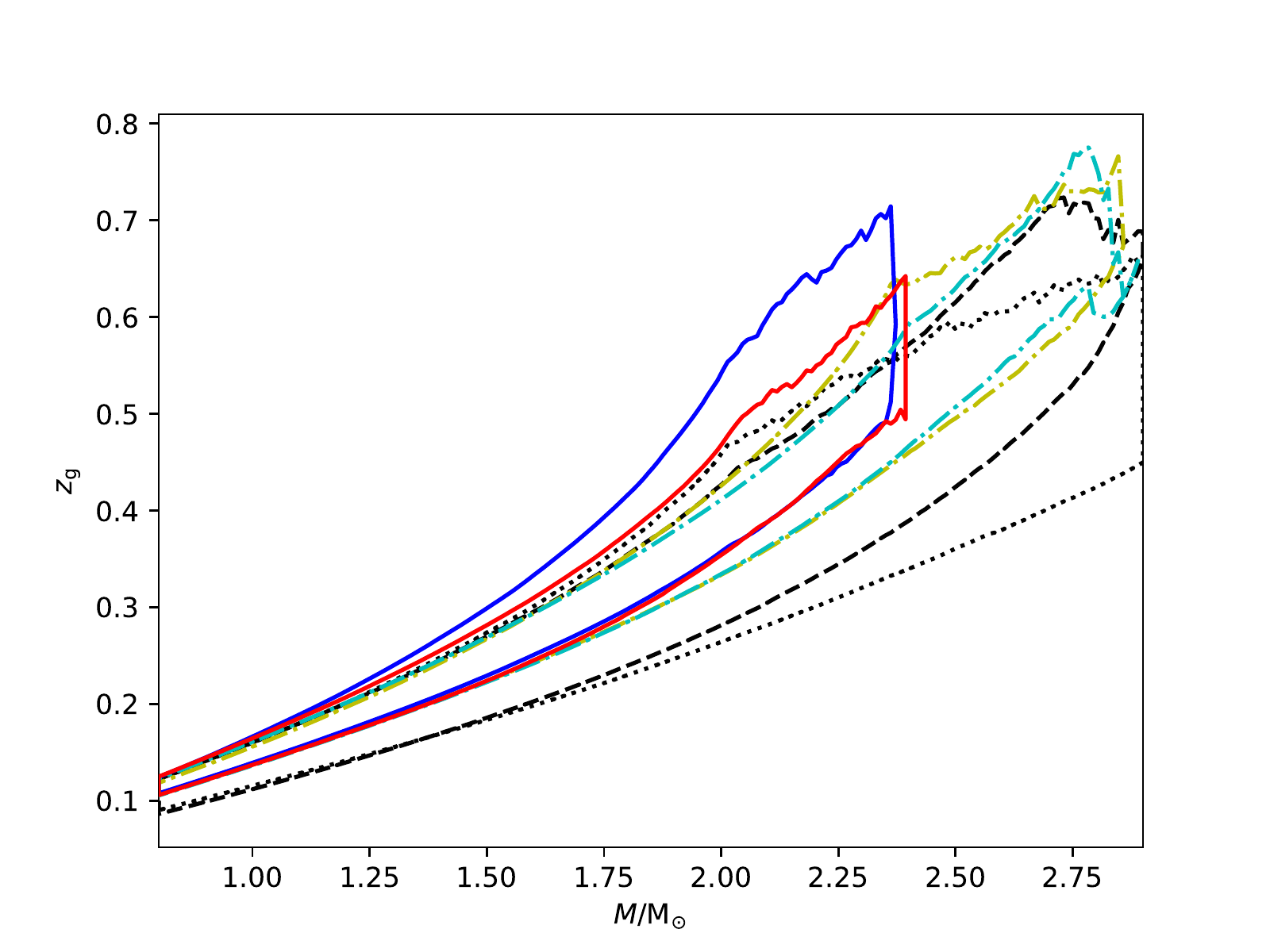}
    \end{minipage}
    }
    \subfigure[]{
    \begin{minipage}[t]{0.47\linewidth}
    \includegraphics[width=3.8in]{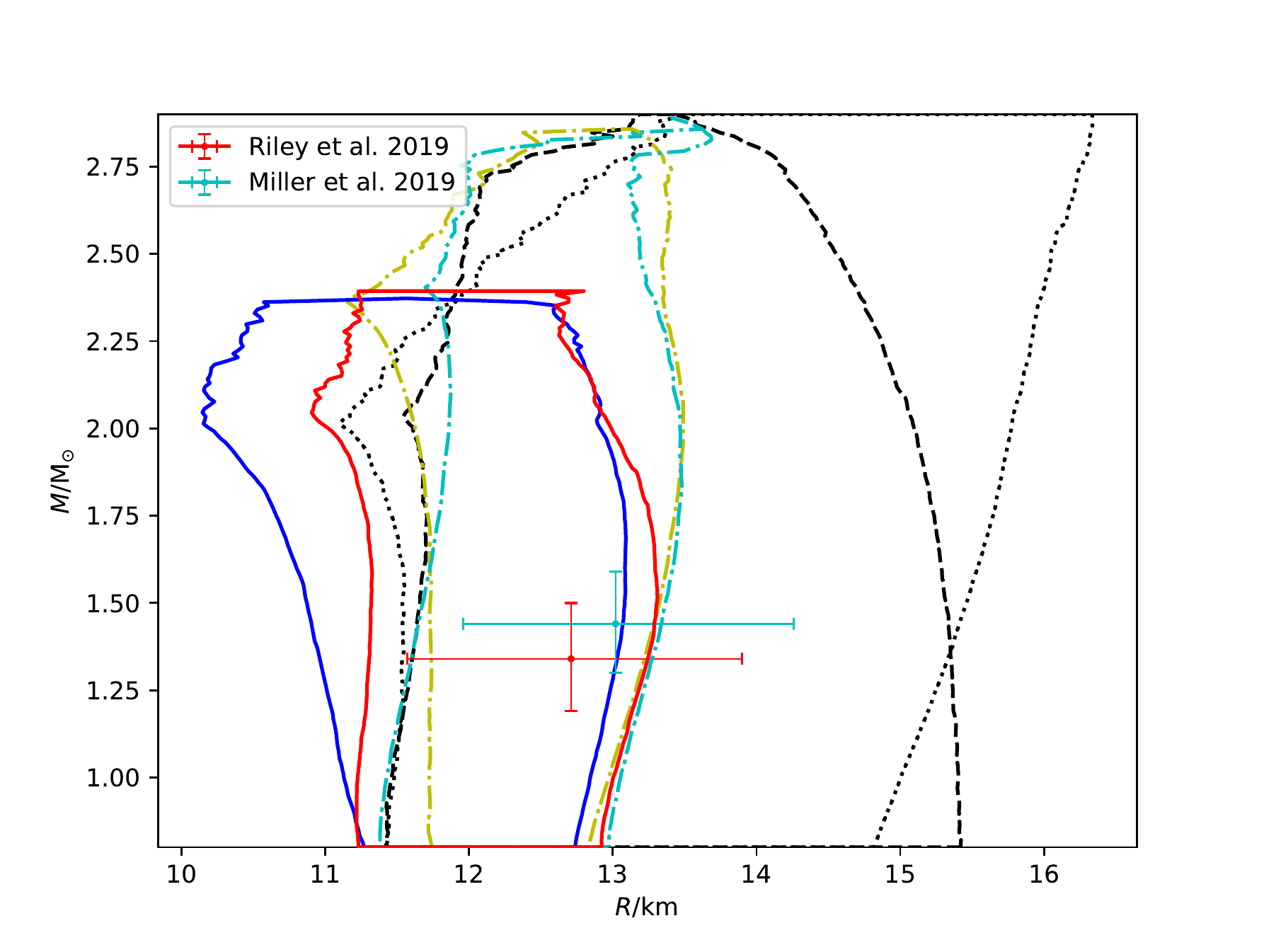}
    \end{minipage}
    }
    \caption{Panels (a)-(d) show the $90\%$ (EoS, sound speed, mass-gravitational redshift, and mass-radius) regions constrained by the data sets described in Section \ref{sec:cons_eos}.  The red (cyan) regions show results from the piecewise method with a constraint from the low (high) $M_{\rm TOV}$ range. The blue (yellow) regions show results from the spectral method with a constraint from the low (high) $M_{\rm TOV}$ range. The black dashed line and the black dotted line show the prior of the piecewise method and the spectral method, respectively. The horizontal purple dashed line in panel (b) represents the limit of $c_{\rm s}/c=1/\sqrt{3}$. The megenta regions in panel (b) represent the $68\%$ and $90\%$ regions of central rest mass density of \objectname{PSR J0030+0451} obtained in the low $M_{\rm TOV}$ case of piecewise analysis. The red and cyan data points in panel (d) are the mass and radius measurement results of \objectname{PSR J0030+0451} obtained from \citet{2019arXiv191205702R} and \citet{2019arXiv191205705M}, respectively.}
\label{fig:eos}
\end{figure}

We construct the $P-\rho$ relation for each posterior sample of $\vec{\theta}_{\rm eos}$ (Fig.\ref{fig:eos}(a)). The narrow pressure uncertainty at $\rho_{\rm sat}$ is mainly governed by the nuclear constraint, and the higher density part between $\rho_{\rm sat}$ and $3\rho_{\rm sat}$ is mainly determined by the gravitational-wave data and the $M-R$ data of \objectname{PSR J0030+0451}, while the highest density region is mainly constrained by the limits on $M_{\rm TOV}$. Meanwhile, for the two investigated $M_{\rm TOV}$ regions, the EoSs at high densities are slightly different, because the more massive of the compact star, the stiffer of the EoS for dense matter.

The character of dense matter can also be described by the sound speed $c_{\rm s}=\sqrt{dp/d\epsilon}$, for which there are two interesting limits $c_{\rm s} \leq c$ and $c_{\rm s} \leq c/\sqrt{3}$ predicted by causality and asymptotically free theories like QCD, respectively. Our results are presented in Fig.\ref{fig:eos}(b), where the megenta region shows the central density of \objectname{PSR J0030+0451}, and the dashed purple line represents the asymptotic limit of $c_{\rm s}=c/\sqrt{3}$. It is unclear whether such a limit has been reached by \objectname{PSR J0030+0451} because of the relatively large uncertainties of both $\rho_{\rm c}$ and $c_{\rm s}$ \citep[see also][]{2019arXiv191205703R, 2019MNRAS.485.5363G}. Future observation of NSs by \emph{NICER} and Advanced LIGO/Virgo are necessary to robustly solve this issue.

The $M-R$ and $M-z_{\rm g}$ relations are also constructed (panels (c) and (d) in Fig.\ref{fig:eos}), and the credible regions are consistent with that of \citet{2019arXiv191205703R}. Interestingly, the measured gravitational redshift of the isolated NS RX J0720.4-3125 \citep[$z_{g}=0.205_{-0.003}^{+0.006}$;][]{2017A&A...601A.108H} is nearly the same as that of \objectname{PSR J0030+0451} \citep[$z_{\rm g} \simeq 0.206_{-0.017}^{+0.014}$;][]{2019arXiv191205702R}. If the NSs share the same EoS and the spin effect is neglected, then the masses of these two isolated objects are expected to be nearly the same, because the radii of NSs are almost unchanged in a relative narrow mass range. For $z_{\rm g}\simeq 0.206$, using results shown in Fig.\ref{fig:eos}(c) we find that the mass of \objectname{PSR J0030+0451} is consistent with that of binary neutron star (BNS) systems \citep{2013ApJ...778...66K}, suggesting no evidence for experiencing significant accretion of these isolated objects \citep[see also][]{2019arXiv191108107T}.

\subsection{Bulk Properties of Neutron Stars}

\begin{figure}[htbp]
    \subfigure[]{
    \begin{minipage}[t]{0.47\linewidth}
    \includegraphics[width=3.8in]{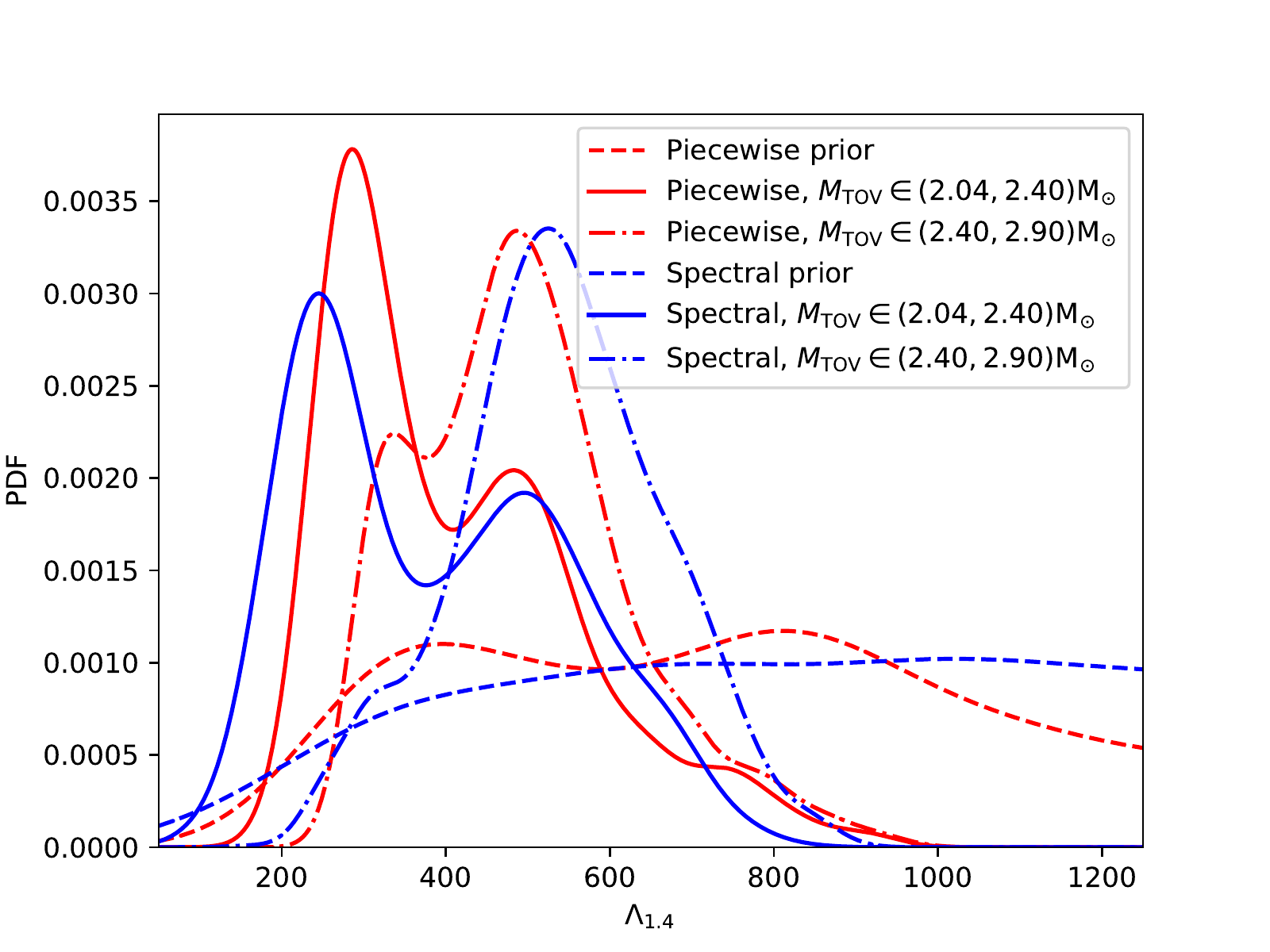}
    \end{minipage}
    }
    \subfigure[]{
    \begin{minipage}[t]{0.47\linewidth}
    \includegraphics[width=3.8in]{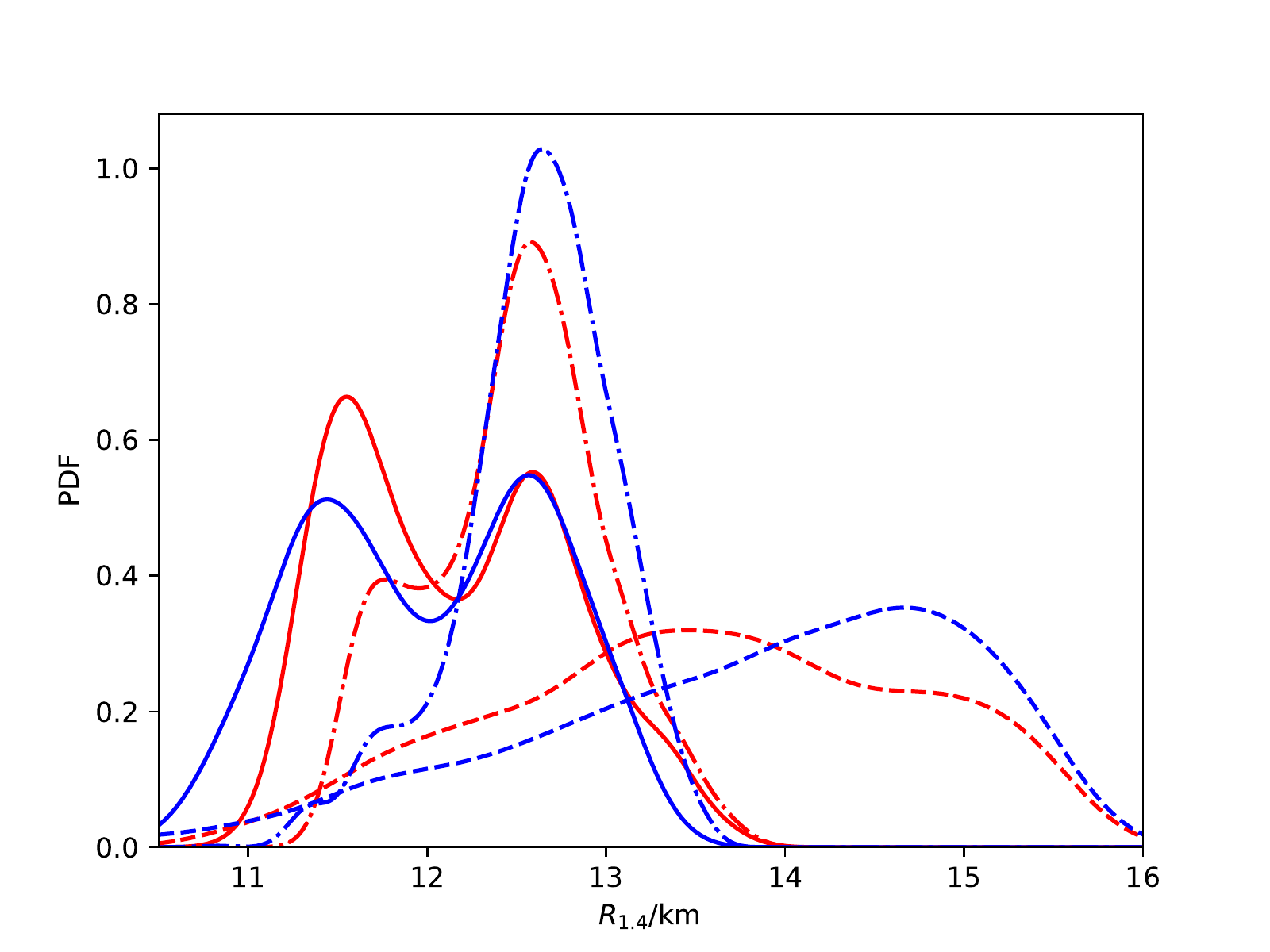}
    \end{minipage}
    }
    \subfigure[]{
    \begin{minipage}[t]{0.47\linewidth}
    \includegraphics[width=3.8in]{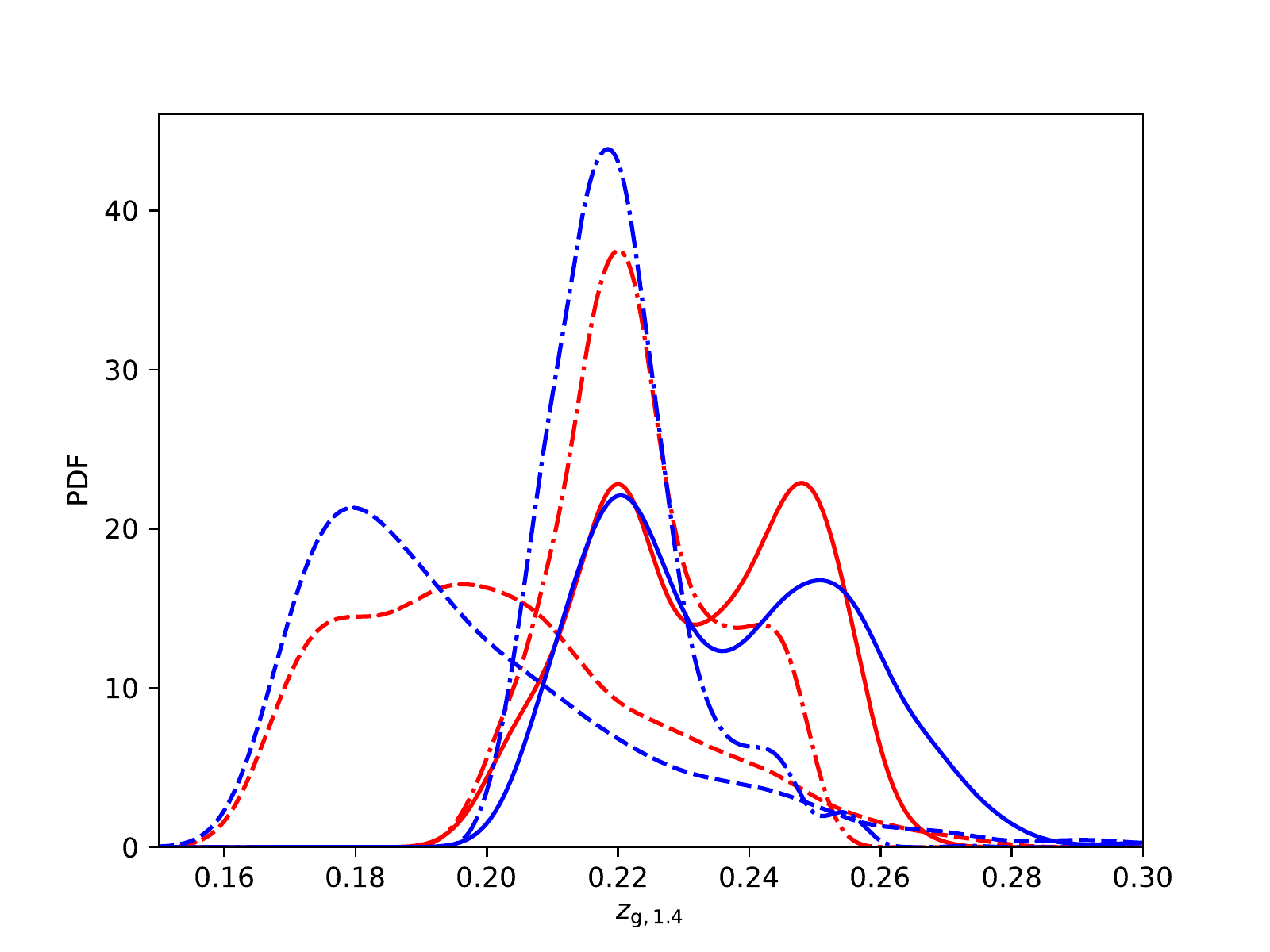}
    \end{minipage}
    }
    \subfigure[]{
    \begin{minipage}[t]{0.47\linewidth}
    \includegraphics[width=3.8in]{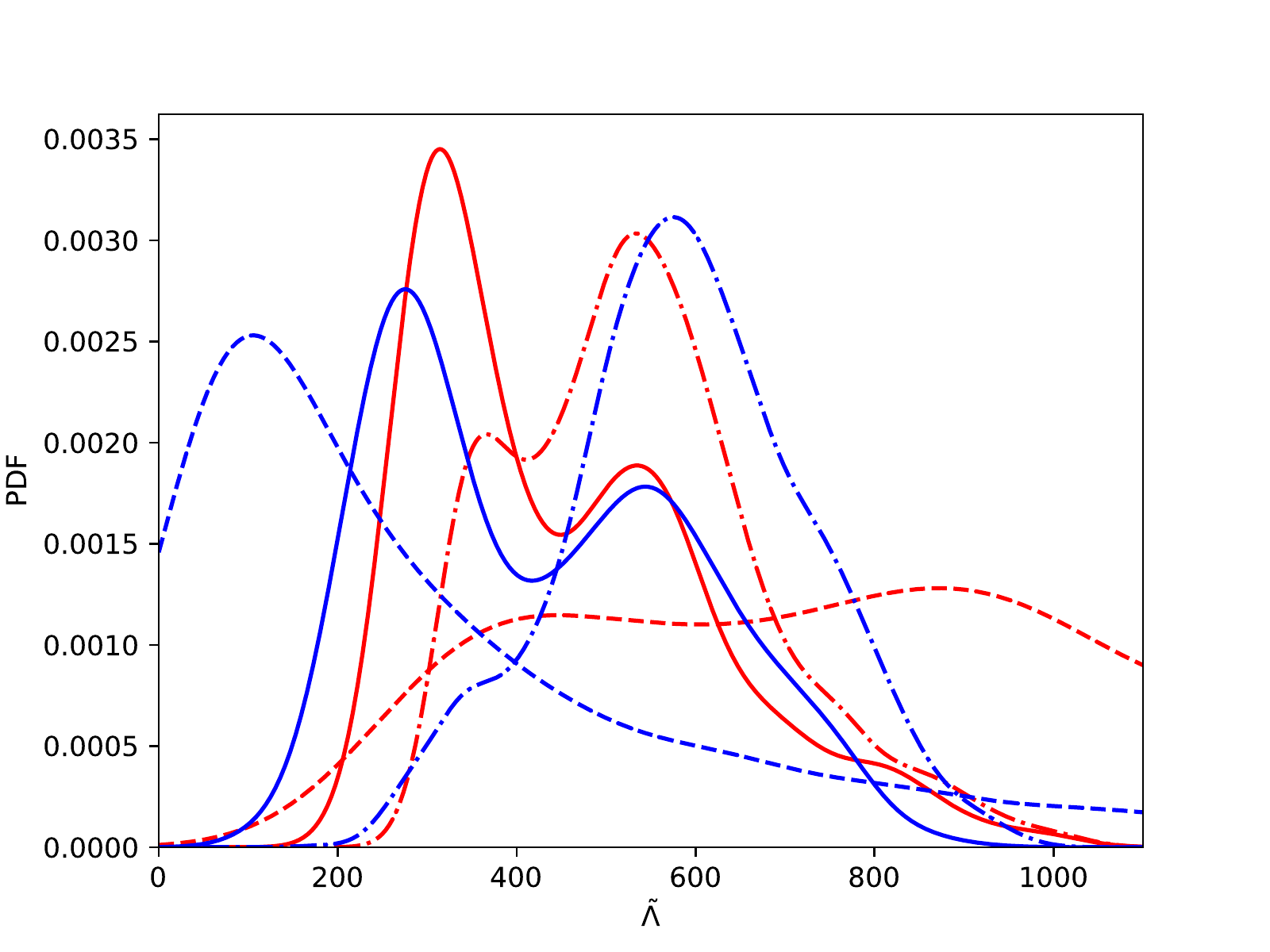}
    \end{minipage}
    }
    \caption{Canonical properties of the 1.4$M_\odot$ NS and $\tilde{\Lambda}$ of GW170817. Panels (a)-(c) show the canonical tidal deformability, radius, and gravitational redshift, respectively. Panel (d) presents the combined dimensionless tidal deformability of GW170817. The red lines and the blue lines show properties obtained from the piecewise method and the spectral method, respectively. The solid, dashed-dotted, and dashed lines represent the cases of $M_{\rm TOV}\in (2.04, 2.40)M_{\odot}$, $M_{\rm TOV}\in (2.40, 2.90) M_{\odot}$, and prior, respectively.}
\label{fig:canonical}
\end{figure}

For a given EoS parameter $\vec{\theta}_{\rm eos}$, the canonical properties of the NSs (or the bulk properties at $M=1.4 M_{\odot}$) can be obtained through optimizing the central pressure $p_{\rm c}$ to reach a gravitational mass $M=1.4 M_{\odot}$. With a group of posterior samples of $\vec{\theta}_{\rm eos}$, we can deduce the probability distributions of bulk properties, such as $\Lambda_{\rm 1.4}$, $R_{\rm 1.4}$, and $z_{\rm g, 1.4}$. As shown in Fig.\ref{fig:canonical}, large $M_{\rm TOV}$ will boost the $\Lambda_{\rm 1.4}$ and $R_{\rm 1.4}$ to higher values in both parameterization methods. This is understandable, because larger $M_{\rm TOV}$ can lead to stiffening of EoS. We also notice that the spectral method is more easily affected by the $M_{\rm TOV}$ condition than the piecewise method, because smooth parameterization models (e.g., spectral method) usually couple the high-density EoS, which sets $M_{\rm TOV}$, from the low-density EoS, that determines $R_{1.4}$ \citep{2019arXiv190810352C, 2019AIPC.2127b0009T}. The canonical gravitational redshift $z_{\rm g, 1.4}$, which is connected to the $R_{\rm 1.4}$ by an one-to-one mapping, mainly reflects the variation contrary to the $R_{\rm 1.4}$. While the combined dimensionless tidal deformability $\tilde{\Lambda}$ of GW170817 simply keeps the same trend as $\Lambda_{\rm 1.4}$. This is mainly because both of the masses of NSs in GW170817 are near the region of $1.4 M_{\odot}$ \citep{2019PhRvX...9a1001A}. Besides, in comparison to using the data of \citet{riley_data}, the incorporation of the data of \citet{miller_data} yields slightly harder EoS in the low $M_{\rm TOV}$ case, while in high $M_{\rm TOV}$ case the results are remarkably consistent.

\begin{table}[]
  \centering
  \begin{ruledtabular}
  \caption{$90\%$ Intervals of Canonical Properties}
  \begin{tabular*}{0.3\textwidth}{lccccc}
  Tests/Properties   &  $R_{\rm 1.4}/\rm km$  & $\Lambda_{\rm 1.4}$  & $z_{\rm g,1.4}$   & $I_{\rm 1.4}/\rm 10^{38} kg \cdot m^2$ & $BE_{\rm 1.4}/ M_{\odot}$    \\
  \hline
    Piecewise prior   &  $13.6_{-1.9}^{+1.8}$  &  $800_{-510}^{+810}$  &  $0.20_{-0.03}^{+0.05}$  &  $1.77_{-0.42}^{+0.37}$  &  $0.14_{-0.02}^{+0.03}$ \\
    Piecewise, Low $M_{\rm TOV}$, RWB\tablenotemark{\dag}  &  $12.1_{-0.8}^{+1.2}$  &  $370_{-130}^{+360}$  &  $0.23_{-0.03}^{+0.02}$  &  $1.43_{-0.13}^{+0.30}$  &  $0.16_{-0.02}^{+0.01}$ \\
    Piecewise, High $M_{\rm TOV}$, RWB  &  $12.5_{-0.9}^{+0.8}$  &  $480_{-180}^{+270}$  &  $0.22_{-0.02}^{+0.02}$  &  $1.54_{-0.17}^{+0.20}$  &  $0.15_{-0.01}^{+0.01}$ \\
    Piecewise, Low $M_{\rm TOV}$, MLD\tablenotemark{\ddag}  &  $12.4_{-1.0}^{+0.8}$  &  $450_{-190}^{+260}$  &  $0.22_{-0.02}^{+0.03}$  &  $1.52_{-0.20}^{+0.20}$  &  $0.16_{-0.01}^{+0.02}$ \\
    Piecewise, High $M_{\rm TOV}$, MLD  &  $12.6_{-0.9}^{+0.8}$  &  $510_{-190}^{+270}$  &  $0.22_{-0.02}^{+0.02}$  &  $1.57_{-0.18}^{+0.19}$  &  $0.15_{-0.01}^{+0.01}$ \\
   \hline
    Spectral prior  &  $14.0_{-2.5}^{+1.4}$  &  $1010_{-730}^{+830}$  &  $0.19_{-0.02}^{+0.06}$  &  $1.90_{-0.54}^{+0.35}$  &  $0.13_{-0.01}^{+0.03}$ \\
    Spectral, Low $M_{\rm TOV}$, RWB  &  $12.0_{-1.1}^{+1.0}$  &  $360_{-180}^{+300}$  &  $0.23_{-0.03}^{+0.03}$  &  $1.42_{-0.22}^{+0.26}$  &  $0.16_{-0.02}^{+0.02}$ \\
    Spectral, High $M_{\rm TOV}$, RWB  &  $12.6_{-0.9}^{+0.6}$  &  $540_{-220}^{+210}$  &  $0.22_{-0.01}^{+0.02}$  &  $1.59_{-0.21}^{+0.15}$  &  $0.15_{-0.01}^{+0.02}$ \\
    Spectral, Low $M_{\rm TOV}$, MLD  &  $12.3_{-1.2}^{+0.8}$  &  $430_{-230}^{+260}$  &  $0.23_{-0.02}^{+0.04}$  &  $1.50_{-0.28}^{+0.20}$  &  $0.16_{-0.01}^{+0.02}$ \\
    Spectral, High $M_{\rm TOV}$, MLD  &  $12.6_{-0.8}^{+0.6}$  &  $530_{-210}^{+210}$  &  $0.22_{-0.01}^{+0.02}$  &  $1.59_{-0.19}^{+0.15}$  &  $0.15_{-0.01}^{+0.01}$ \\
  \end{tabular*}
  \tablenotetext{\dag}{Posterior sample of \citet{riley_data} used}
  \tablenotetext{\ddag}{Posterior sample of \citet{miller_data} used}
  \label{tb:canonical}
  \end{ruledtabular}
\end{table}

\begin{figure}[htbp]
    \subfigure[]{
    \begin{minipage}[t]{0.3\linewidth}
    \includegraphics[width=2.5in]{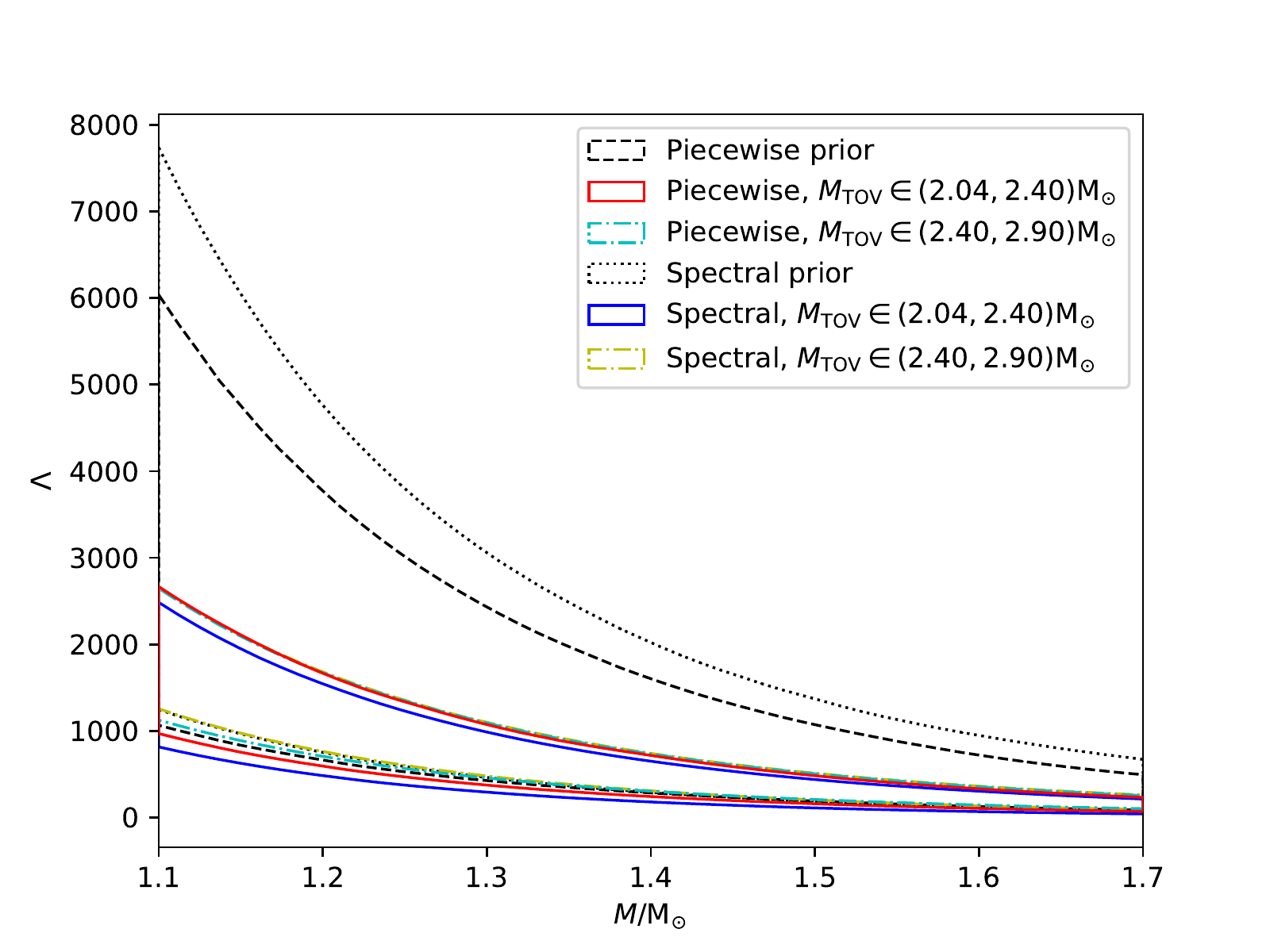}
    \end{minipage}
    }
    \subfigure[]{
    \begin{minipage}[t]{0.3\linewidth}
    \includegraphics[width=2.5in]{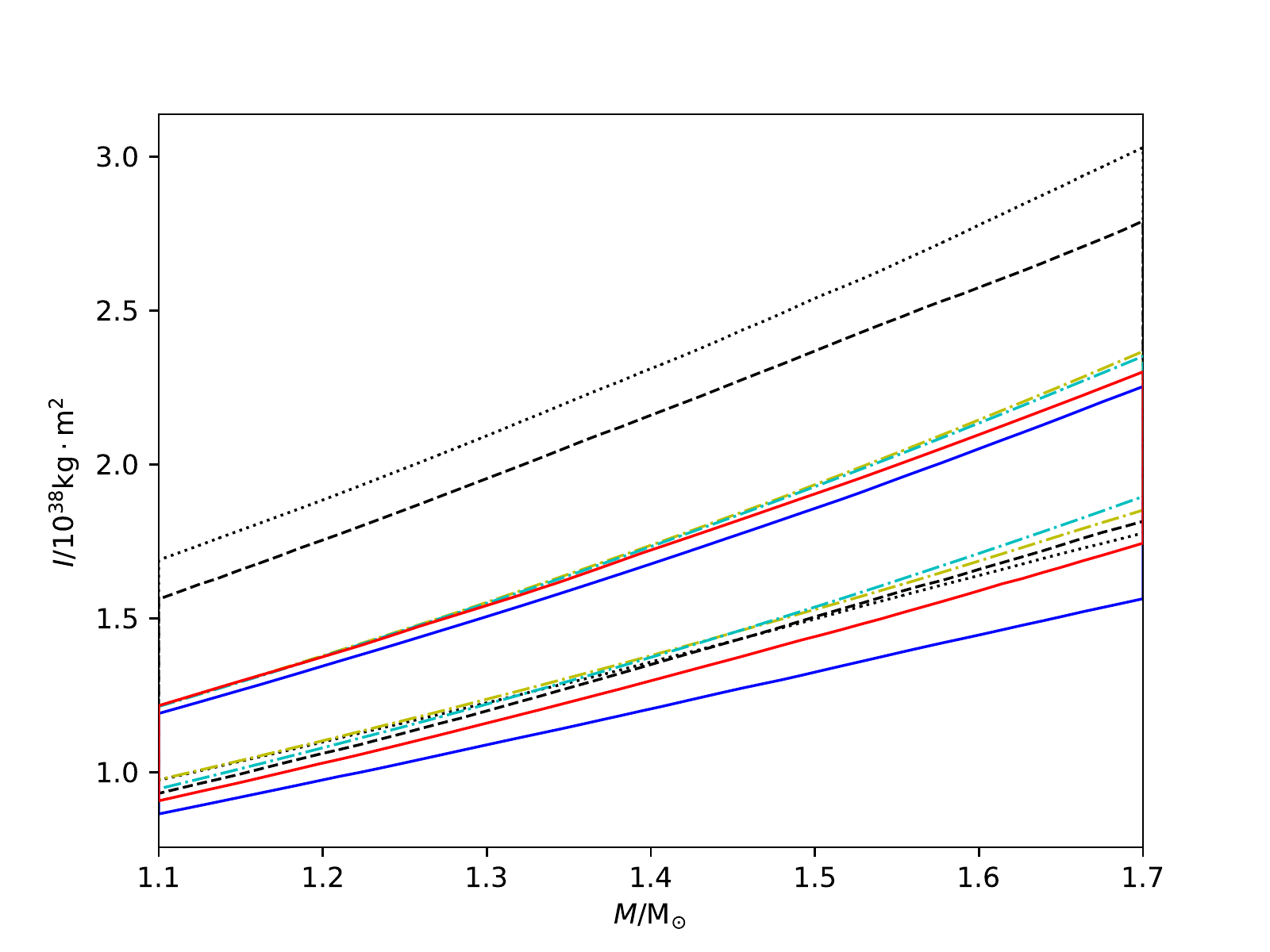}
    \end{minipage}
    }
    \subfigure[]{
    \begin{minipage}[t]{0.3\linewidth}
    \includegraphics[width=2.5in]{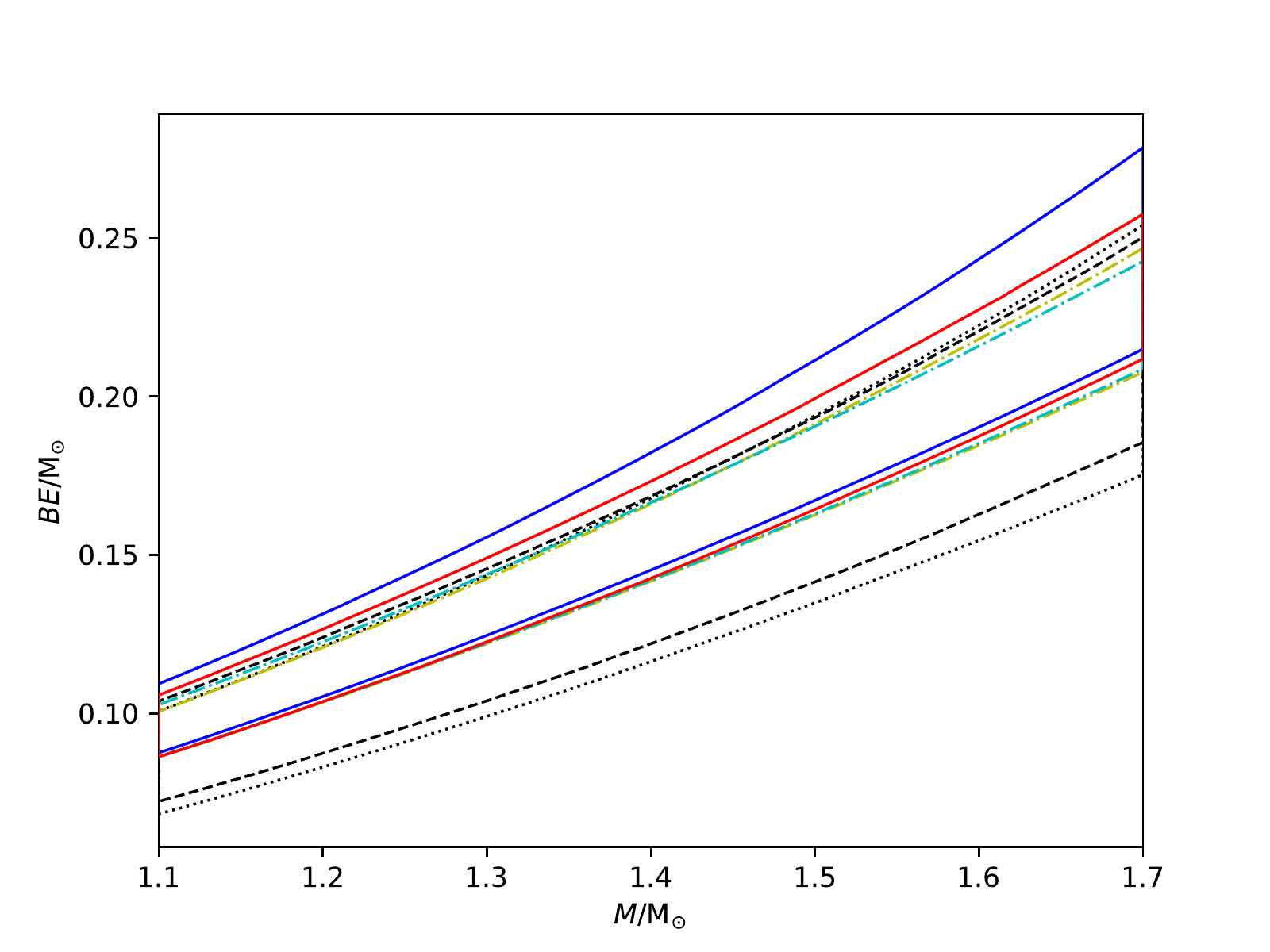}
    \end{minipage}
    }
    \caption{Same as Fig.\ref{fig:eos} but for bulk properties of NSs inferred from universal relations. Panel (a), (b), and (c) show the possible regions of respectively tidal deformability, moment of inertia, and binding energy in the mass range of $(1.1-1.7) \rm M_{\odot}$.}
\label{fig:bulk}
\end{figure}

With the posterior samples of three pairs of $M-\Lambda$, namely, $\{M_1, \Lambda_1, M_2, \Lambda_2, M_3, \Lambda_3 \}$, it is possible to deduce some bulk properties of NSs, as done in \citet{2018ApJ...868L..22L,2018PhRvL.121p1101A}. Here we adopt three universal relations to transform our posterior samples into constraints of tidal deformability, moment of inertia, and binding energy of NSs in the mass range $(1.1-1.7) M_{\odot}$. These relations \citep[see][and reference therein]{2017PhR...681....1Y} read
\begin{equation}
    \label{eq:BE}
    \begin{aligned}
    \lambda{(M)} & \simeq \lambda_{\rm ref} + \lambda^{1}(M-M_{\rm ref})/M_{\odot}, \\
    \log_{10}{\bar{I}} & = \sum_{n=0}^{4}a_{\rm n}(\log_{10}{\Lambda})^n,  \\
    BE/M & = \sum_{n=0}^{4} b_{\rm n}\bar{I}^{-n},
    \end{aligned}
\end{equation}
where $\lambda(M) \equiv \Lambda(M) (GM/c^2)^5$ is the tidal deformability (its dimensionless form is $\Lambda$), $\bar{I} \equiv c^4I/G^2M^3$ is the dimensionless moment of inertia (its dimensional form is $I$), $BE$ is the binding energy, and $G$ is the Newton's gravitational constant. We take the coefficients $a_{\rm n}$ and $b_{\rm n}$ from \citet{2018ApJ...868L..22L} and \citet{2016EPJA...52...18S}, respectively. For each reference mass $M_{\rm ref}$, we can get a best fit of $\lambda_{\rm ref}$ and $\lambda^{1}$ with a single posterior sample $\{M_1, \Lambda_1, M_2, \Lambda_2, M_3, \Lambda_3 \}$. Repeating this process for a population of posterior samples, we obtain a sample of $\lambda_{\rm ref}$ and $\lambda^{1}$. The $\lambda_{\rm ref}$ is inserted into the $\bar{I}-\Lambda$ relation to get $\bar{I}$. Then, the resulting samples of $\bar{I}$ are inserted into the $BE-\bar{I}$ relation to obtain samples of $BE$. We finally get constraints on $\Lambda$, $I$, and $BE$ for each reference mass in the range of $(1.1-1.7) M_{\odot}$. We find that all the four tests give results that are consistent with each other and shrink the prior of EoS to a softer region, with high $M_{\rm TOV}$ cases showing slightly larger $\Lambda$, larger $I$, and lower $BE$ due to the stiffening of EoS in these cases (as shown in Fig.\ref{fig:eos}(a) and Table \ref{tb:canonical}). For \objectname{PSR J0737-3039A}, which has an accurately measured mass of $1.338 M_\odot$ and is expected to have a precise determination of moment of inertia in the next few years via radio observations, our Test (i) predicts its moment of inertia $I={1.35}^{+0.26}_{-0.14} \times \rm 10^{38} kg \cdot m^2$. This value is higher than that of \citet{2018ApJ...868L..22L}, which is mainly caused by the fact that they adopted a group of $\Lambda_{\rm 1.4}$ smaller than ours.

\section{Conclusion and Discussion}
\label{sec:conc_disc}

In this work, we perform parameter estimation of EoS using two kinds of parameterization methods, with the information obtained from the combination of latest $M-R$ measurements of \objectname{PSR J0030+0451} from \emph{NICER}, strain data of GW170817, and constraints from nuclear experiments/theories. Our results show that, with the additional inclusion of the robust measurements of $M-R$ of \objectname{PSR J0030+0451}, the uncertainty region of the $P-\rho$ diagram is reduced compared to previous works. \citet{2016EPJA...52...18S} and \citet{2015PhRvL.114c1103B} have shown that the sound velocity at ultra-dense matter can exceed $c/\sqrt{3}$, considering the nuclear theories and the maximum observed NS mass. As for \objectname{PSR J0030+0451}, the situation is unclear because of the relatively large uncertainties of both $\rho_{\rm c}$ and $c_{\rm s}$ (see Fig.\ref{fig:eos}(c)).

Meanwhile, bulk properties of NS, e.g., $\Lambda_{1.4}$, $R_{1.4}$, have been better determined. For \objectname{PSR J0737-3039A} with the mass of $1.338 M_\odot$ we predict a moment of inertia $I={1.35}^{+0.26}_{-0.14} \times \rm 10^{38} kg \cdot m^2$. Using the $z_{\rm g}-M$ relation, the isolated NS RX J0720.4-3125 \citep[$z_{g}=0.205_{-0.003}^{+0.006}$;][]{2017A&A...601A.108H} is evaluated to have nearly the same mass as \objectname{PSR J0030+0451}, which favors BNS mass distribution \citep{2013ApJ...778...66K}, consistent with our previous work \citep{2019arXiv191108107T}. We also notice the difference between the two parameterization methods and the impact of different choices of $M_{\rm TOV}$ ranges. For both methods, higher $M_{\rm TOV}$ presents the stiffening of EoS above $2 \rho_{\rm sat}$, along with larger $R_{1.4}$ and $\Lambda_{1.4}$. Besides, smooth EoS models (e.g., spectral method), which couple the EoS in the whole ranges of density, are more sensitive to $M_{\rm TOV}$ constraints \citep{2019arXiv190810352C, 2019AIPC.2127b0009T}. Though the results are consistent with each other if the uncertainties have been taken into account, these phenomena caution that a reliable region of $M_{\rm TOV}$ and the choice of parameterizing methods are important for constraining the EoS.

With the accumulated observation data of \emph{NICER}, the $M-R$ measurements of more targeted NSs \citep{2019arXiv191205708G}, such as \objectname{PSR J1614-2230} and \objectname{PSR J0740+6620}, will be obtained with an unprecedented accuracy thanks to the remarkable performance of \emph{NICER}. Therefore, the EoS can be effectively determined with the radii measurements of NSs with various masses as shown in \citet{2019ApJ...881...73W}. Moreover, reducing the uncertainty of radius from $10\%$ to $5\%$, \citet{2018A&A...616A.105S} showed that $\sim 10\%$ and $\sim 40\%$ accuracy in central parameter estimation can be obtained for low-mass and high-mass NSs, respectively. Thus, the behavior of sound velocity at ultra-dense matter and whether the phase transition occurs in NS can be reliably probed. Meanwhile, with the upgrade of Advanced LIGO/Virgo detectors, more and more BNS and neutron star black hole (NSBH) merger events will be caught. For instance, quite a few BNS/NSBH candidates have been reported in the LIGO/Virgo O3 public alerts (GraceDB\footnote{\url{https://gracedb.ligo.org/superevents/public/O3/}}). A sample of $\sim 100$ BNS merger events is expected to tightly constraint the pressure of neutron star matter in a wide density region \citep{arXiv:1904.04233}. Therefore, it is feasible to robustly determine the EoS with the gravitational-wave data and the $M-R$ measurements from \emph{NICER} in the near future.

\acknowledgments
We thank the anonymous referee for the helpful suggestions. This work was supported in part by NSFC under grants of No. 11525313 (i.e., Funds for Distinguished Young Scholars) and No. 11921003, the Chinese Academy of Sciences via the Strategic Priority Research Program (grant No. XDB23040000), Key Research Program of Frontier Sciences (No. QYZDJ-SSW-SYS024). This research has made use of data and software obtained from the Gravitational Wave Open Science Center \url{https://www.gw-openscience.org}, a service of LIGO Laboratory, the LIGO Scientific Collaboration and the Virgo Collaboration. LIGO is funded by the U.S. National Science Foundation. Virgo is funded by the French Centre National de Recherche Scientifique (CNRS), the Italian Istituto Nazionale della Fisica Nucleare (INFN) and the Dutch Nikhef, with contributions by Polish and Hungarian institutes.

\software{Bilby \citep[version 0.5.5, ascl:1901.011, \url{https://git.ligo.org/lscsoft/bilby/}]{2019ascl.soft01011A}, LALSuite \citep[version 6.57, \url{https://git.ligo.org/lscsoft/lalsuite}]{lalsuite}, PyCBC \citep[version 1.13.6, ascl:1805.030, \url{https://pycbc.org}]{2018ascl.soft05030T}, PyMultiNest \citep[version 2.6, ascl:1606.005, \url{https://github.com/JohannesBuchner/PyMultiNest}]{2016ascl.soft06005B}.}

~

~

~

\bibliographystyle{aasjournal}

\end{document}